\documentclass[journal]{IEEEtran}
\UseRawInputEncoding
\usepackage[utf8]{inputenc}
\usepackage{amsfonts,amsmath, amsthm,bbm}
\usepackage{subcaption}
\usepackage{algorithm}
\usepackage{array}

\usepackage{textcomp}
\usepackage{hyperref}
\hypersetup{
    colorlinks=true,
    linkcolor=cyan,
    filecolor=magenta,      
    urlcolor=blue,
    pdftitle={Overleaf Example},
    pdfpagemode=FullScreen,
    }
    
\usepackage{stfloats}
\usepackage{url}
\usepackage{multirow}
\usepackage{verbatim}
\usepackage{graphicx}
\usepackage{tabularx} 
\usepackage{algpseudocode}
\usepackage{soul}
\usepackage[dvipsnames]{xcolor}
\usepackage{enumitem}
\usepackage{bm}
\usepackage{booktabs} 
\usepackage{cite}
\usepackage{orcidlink}

\newcommand{\capoptix}{\textsc{CapOptix}}
\newtheorem{definition}{Definition}
\newtheorem{theorem}{Theorem}
\newtheorem{proposition}{Proposition}
\newtheorem{remark}{Remark}

\newtheorem{lemma}{Lemma}
\newtheorem{assumption}{Assumption}

\begin{document}

\title{\capoptix: An Options-Framework for Capacity Market Pricing}

\author{
Millend Roy \orcidlink{0000-0003-3054-954X},
 Agostino Capponi \orcidlink{0000-0001-9735-7935},
 Vladimir Pyltsov \orcidlink{0009-0007-2580-3853},
 Yinbo Hu \orcidlink{0000-0001-7191-6743}, and Vijay Modi \orcidlink{0000-0003-2513-0437} 
\IEEEmembership{}

\thanks{Millend Roy and Agostino Capponi are with the Department of
Industrial Engineering and Operations Research, Columbia University, New York 10027, New York, USA (e-mail: millend.roy@columbia.edu; ac3827@columbia.edu). Agostino Capponi holds a courtesy appointment at the Columbia Business School.}
\thanks{Vladimir Pyltsov, Yinbo Hu and Vijay Modi are with the Department of Mechanical Engineering, Columbia University,  New York 10027, New York, USA (e-mail:  v.pyltsov@columbia.edu, yh3184@columbia.edu, modi@columbia.edu).}
}

\markboth{Journal of \LaTeX\ Class Files,~Vol.~14, No.~8, December~2025}%
{Shell \MakeLowercase{\textit{et al.}}: A Sample Article Using IEEEtran.cls for IEEE Journals}

\IEEEpubid{0000--0000/00\$00.00~\copyright~2025 IEEE}

\maketitle

\begin{abstract}
Electricity markets are under increasing pressure to maintain reliability amidst rising renewable penetration, demand variability, and occasional price shocks. Traditional capacity market designs often fall short in addressing this by relying on expected-value metrics of energy unserved, which overlook risk exposure in such systems. In this work, we present \capoptix, a capacity pricing framework that interprets capacity commitments as reliability options, i.e., financial derivatives of wholesale electricity prices. 
\capoptix \text{} characterizes the capacity premia charged by accounting for structural price shifts modeled by the Markov Regime Switching Process. We apply the framework to historical price data from multiple electricity markets and compare the resulting premium ranges with existing capacity remuneration mechanisms.
\end{abstract}


\section{Introduction}
\label{sec:introduction}
\IEEEpubidadjcol
\IEEEPARstart{T}{he} global energy transition toward renewable and low-carbon generation introduces fundamental challenges for electricity markets and system operators \cite{kirschen2018fundamentals}. As reliance on intermittent energy sources increases, so does the frequency and intensity of demand-supply imbalances, commonly referred to as security of supply problems \cite{creti2019economics}. These imbalances arising not only from renewable intermittency \cite{IEMRE} but also from extreme weather events, transmission congestion, unplanned outages, or scheduled maintenance, can result in periods of system stress where available generation falls short of demand. 

In an energy-only market, generators are compensated solely based on the spot price of electricity, with scarcity-driven price spikes expected to incentivize investment \cite{ conejo2016investment}. However, this mechanism fails in practice, leading to what is widely recognized as the missing money problem \cite{stoft2002power}, elevating revenue uncertainty \cite{morales2013integrating}. This growing disconnect between the system’s long-term generation adequacy needs and short-term market incentives highlights the necessity of complementary mechanisms like the Capacity Remuneration Mechanism (CRM), which is a set of policy tools explicitly designed to remunerate generators for maintaining available capacity, even when not actively dispatched \cite{creti2019economics}.


\IEEEpubidadjcol 
Several types of CRMs \cite{kirschen2018fundamentals} have been adopted across electricity markets which include: \textit{capacity payments}, which offer administratively determined compensation for availability (e.g., Portugal \cite{CEER2019_Portugal}, Spain \cite{IEEFA2016_SpainCapacity}); \textit{capacity obligations}, which require load-serving entities to contract sufficient capacity to meet expected load (e.g., Midcontinent Independent System Operator (MISO) \cite{MISO2023_BPM011}, California Independent System Operator (CAISO) \cite{CAISO2024_Section43A}); \textit{strategic reserves}, where specific plants are withdrawn from the market and retained for emergency use (e.g., Sweden \cite{Svk2023_CapacityMechanism}, Germany \cite{creti2019economics}); and \textit{capacity auctions}, which rely on market-based price discovery to procure capacity for future delivery (e.g., Pennsylvania, New Jersey, Maryland Interconnection (PJM) \cite{PJMFactSheet2025}, New York Independent System Operator (NYISO) \cite{nyiso_capacity_market}). A notable variant of the auction approach \cite{capacity_market_carmton} is the \textit{reliability option}, implemented in regions like ISO New England \cite{ ISO_NE_IMMU_2008}, Italy \cite{RSE_CapacityMarket_2024}.

\IEEEpubidadjcol 
While these mechanisms reflect diverse regulatory preferences and market conditions, a unified, risk-sensitive framework for pricing capacity efficiently remains absent. Many suffer from inefficient price signals, rigid administrative rules, or poor alignment with investment risk. For years, the Federal Energy Regulatory Commission (FERC) and some Regional Transmission Organizations (RTOs) have said capacity-market prices are too low \cite{AAGAARD2022101335}. Paradoxically, under the current market design principles, the evidence points the other way \cite{AAGAARD2022101335}: ``These markets buy more capacity than what reliability actually needs."
Because the demand curve is set by administrators, it leads to systematic over-procurement, and consumers end up paying billions for extra capacity they don’t use regularly. Bhagwat et al. \cite{bhagwat2017effectiveness}'s agent-based NYISO Installed Capacity market model, Chattopadhyay et al. \cite{chattopadhyay2015capacity}'s Cournot with transmission, Hach et al. \cite{hach2016capacity}'s dynamic investment, and Bothwell et al. \cite{bothwell2017crediting}'s equilibrium crediting models improve price formation and investment insights but rely on expected values of energy unserved, overlooking extreme scarcity events that dominate high-renewable grids. These events are better captured using jump processes, which are rarely applied to capacity pricing.\footnote{A notable exception is Andreis et al. \cite{ANDREIS2020104705}, who derive closed-form reliability option premia, but they ignore regime shifts, scarcity-hour illiquidity, and structural market volatility. Jump-diffusion models have instead been widely used for spot price risk management \cite{pricing_mean_reverting}.}     


Among the available designs, reliability options (RO) stand out as a market-compatible approach. These are derivative-like contracts that pay generators a premium to supply energy at a capped strike price during scarcity periods. The core idea is to \textit{identify all potential events where supply shortfalls and corresponding price spikes may occur, and to compute the aggregate discounted financial impact ahead of time}. This aggregated-discounted impact is precisely the compensation, or capacity premium, demanded by generators to supply electricity under these circumstances.

In practice, the premium is discovered through an auction in which generators bid for RO contracts. Motivated by the option-like payoff structure of these contracts, we develop \capoptix, a pricing framework that formalizes the value of the capacity premium using the theory of financial options.\footnote{Auction formats (e.g., descending-clock or sealed-bid auctions) determine how generators’ valuations are translated into bids and clearing prices. Our focus in this paper is on the underlying economic value of the reliability option as a contingent claim on scarcity prices, not on strategic bidding behavior. An explicit auction-design analysis would be complementary to, rather than a substitute for, the derivative-based valuation framework we propose.}

We simulate a range of possible future price paths to evaluate the expected payoff of the reliability option, thereby deriving the capacity premium as the fair value of that option. Additionally, \capoptix \text{} prescribes suitable values for its design parameters, i.e., (i) \textit{Strike Price through tail-risk awareness}: We prove that the inherent option structure aligns with tail-risk-aware metrics such as Conditional Value-at-Risk (CVaR), enabling strike price selection to be the system’s $\alpha$-quantile risk set by the operator (typically, known as option-based portfolio insurance (OBPI) \cite{PeroldSharpe1995}). 
(ii) \textit{Minimum Contract Duration by cost-revenue break-even}: We show that the resulting capacity premia converges to the Net Cost of New Entry (NetCONE)\footnote{Check \href{https://www.pjm.com/-/media/committees-groups/committees/mrc/2023/20230920/20230920-item-06---1-local-considerations-of-net-cone---presentation.ashx}{here} and Section \ref{sec:contractduration} for NetCONE details.} (a critical benchmark to assess provision of sufficient investment incentives), determining minimum contract durations for investment viability.

We validate the generality and effectiveness of the approach by conducting empirical simulations using real-world data from New York (NYISO), California (CAISO), Texas (ERCOT), Germany, and Italy, each characterized by different levels of price volatility, renewable penetration, and regulatory design. To account for this heterogeneity, \capoptix \text{}  explores a suite of stochastic processes to model electricity prices, including (i) a mean-reverting process with Poisson jumps, (ii) Generalized Autoregressive Conditional Heteroskedasticity (GARCH) processes \cite{bollerslev1986generalized} with jumps, and (iii) Markov regime-switching models with a mean-reverting behavior that can capture both long-term structural shifts and sudden, rare events such as market shocks or extreme weather-induced supply disruptions. 
By tailoring the price dynamics to each region’s empirical characteristics, we ensure that the derived capacity premia accurately reflect the local market conditions.



\section{Modeling Assumptions and Specifications}
\label{sec:assumptions}
We use option pricing theory from finance to accurately estimate capacity premia. In this section, we first list the assumptions for our problem formulation and then dive into the details on how to model it.


\begin{assumption}[Price-Taking Behavior]
     Upon constructing the generation portfolio, the generator retains the right to participate in wholesale electricity markets or engage in forward contracts. The generator works as a price taker, i.e., it can respond to market prices but does not influence them.
\end{assumption}

\begin{assumption}[Constant O\&M Rate]\label{as:const_OandM}
We levelize the operating-and-maintenance expense per unit of installed capacity and assume it is constant over the horizon \([0,T+\tau]\), $\text{O\&M}(t)\;=\;\text{O\&M}$ for $0\le t\le T+\tau$.
\end{assumption}

For a price-taking capacity provider, especially for a renewable portfolio player, \(\text{O\&M}\) is typically an order of magnitude smaller than the market-clearing energy price; consequently, modeling time-varying O\&M has a negligible effect on the revenue-cost balance and is omitted for tractability.

\begin{assumption}[Scarcity–pricing]\label{ass:scarcity}
For every delivery time $t$ the energy price $S_t$ satisfies: $S_t \;=\; s_{\text{base}_t} \;+\; \varphi\!\bigl(Z_t(\mathcal{R})\bigr)$,
where $s_{\text{base}_t}$ is the energy-only price that is last bid price by the marginal generator, $Z_t(\mathcal{R})$ is the \emph{instantaneous generation shortfall i.e. difference between the highest supply bid volume to the lowest demand bid volume}\footnote{In reality, it is difficult to analyze shortfall from real-world data since the logged data includes imports to meet the excess demand or demand gets curtailed when not met. Note that $S_t$ is generally the price of short-term markets like the day-ahead market or real-time market. At times of excess demand, the supply and demand bids do not intersect, leading to extrapolation or shifting of curves to find the non-materialized clearing point. 
}, and $\varphi:\mathbb R_+\to\mathbb R_+$ is a shortfall to price map, where $\varphi$ is strictly increasing for $z>0$. 
\end{assumption}

\noindent Having specified the assumptions, we now present the model. 

\begin{enumerate}
\item At contract initiation ($t=0$), the capacity provider receives an upfront premium from the load-serving entities (LSEs). Over the preparation horizon $[0,T]$, the provider undertakes capacity planning and investment to ensure readiness for future delivery obligations.

\item Throughout $[0,T+\tau]$, the provider may participate in wholesale, bilateral, and futures markets to finance infrastructure and manage capital risk.

\item During the delivery period $[T, T+\tau]$, the LSEs hold the option without obligation to procure electricity at the strike price $K_t$ whenever market prices exceed it. In such cases, the provider must supply power at $K_t$ or refund the difference $(S_t - K_t)$, ensuring the LSE is fully hedged against scarcity price spikes. 
\end{enumerate}




Capacity market contracts commit to clear a fixed capacity $Q$ for delivery at each interval $\tau_i$. To draw an analogy, electricity delivered during times t=$[T,T+\tau]$, within the capacity limits of Q, is akin to holding a portfolio of $\tau$ options that can be exercised at different maturity dates spanning over T to $T+\tau$. This is equivalent to a strip of $\tau$ European Call Options maturing on successive time intervals from $T$ to $T+\tau$, each exercisable if market conditions justify.

This setup enables a tractable formulation for pricing the capacity premium based on classical financial option theory.



\begin{proposition}[Capacity Premium as a Strip of European Options]
\label{ROasEO}
Let \( \{S_t\}_{t \geq 0} \) denote the stochastic process representing the wholesale electricity spot price. Suppose a capacity contract provides a fixed capacity \( Q \) over the delivery window \( [T, T+\tau] \), where \( T \) is the infrastructure maturity time and \( \tau \) is the delivery period. The strike price is predetermined as $K_t$\footnote{While in practice, the strike price \( K_t \) varies across delivery intervals \cite{ANDREIS2020104705} due to changes in raw material costs, marginal plant efficiency, or market conditions. Since such fluctuations follow a typical mean-reverting behavior, there exists no persistent drift, and the expected price level remains "stable" over time. To ensure tractability and to preserve contract attractiveness for LSEs, who value price certainty, while computing our results, we assume a constant \( K \) throughout the contract period. Frequent variations in \(K_t\) could erode LSE confidence and deter from option contract participation similar to the problem of valuation uncertainty in OTC options \cite{duffie2011dark}.  
}. 
Then, the capacity premium \( C(0, S_0) \) charged per unit of contracted capacity $Q$ at the initial time \( t = 0 \) is given by:
\begin{align}
    C(0, S_0) = \sum_{t = T}^{T + \tau} e^{-rt} \, \mathbb{E} \left[ \left( S_t - K_t \right)^+ \right] \Delta t
\end{align}
where $[\cdot]^+ = \max(\cdot,0)$, \( r \) is the risk-free discount and \( \Delta t \) is the sample time of delivery (e.g., 15 mins or hourly).
\end{proposition}

\noindent Here, we calculate $\mathbb{E}[\cdot]$ using the risk-neutral pricing measure\footnote{ISO/RTO prices are realized physical prices and are naturally modeled under the historical measure \(\mathbb{P}\). In contrast, power forwards and futures traded on financial exchanges (e.g., EEX, CME) are priced for hedging and are approximated by risk-neutral expectations. For our empirical results, we work directly under  \(\mathbb{P}\), using realized ISO prices to compute RO premia.}. 

\section{Contract Duration for new Capacities}
\label{sec:contractduration}
In this section, we estimate one of the key parameters of the option pricing model, i.e., the contract duration $\tau$, which directly influences the premium value. 

We start by defining Net Cost of New Entry ($NetCONE$) by accounting explicitly for the revenues that a newly constructed generating resource expects to earn from energy and ancillary services (AS) markets.
\begin{definition}[NetCONE]
    \label{def:netcone}
  NetCONE represents the residual annual revenue required beyond market-based earnings from energy and ancillary services (AS) to ensure new investments in generation capacity are financially viable.
 \[
 NetCONE = CONE - \mathbb{E}[\text{Revenue from }(Energy + AS)]
 \]
 where $CONE$ (Cost of New Entry) is the annualized total fixed cost (capital investment, fixed Operations and Management (O\&M), financing) required to build and operate a new generating resource per MW-year.
\end{definition}

However, under Reliability Options (RO), capacity providers forgo revenues from energy markets above a strike price $K$ and are compensated from the upfront $RO$ premium. Consequently, under the RO structure, the premium closely aligns with $NetCONE$. 

\begin{proposition}
\label{premium_as_netcone}
    Under the reliability options framework, the optimal capacity remuneration, i.e., the reliability option premium ($C$), is equal to the $NetCONE$.
\end{proposition}

For investment viability, the total expected present value of revenues must equal or exceed the total present value of costs, i.e., $PV(Total Revenue) \geq PV(Total Cost)$.
From a break-even perspective, the above equation becomes : 
    \begin{align}
    \label{eq:breakeven}
            PV(Total Revenue) = PV(Total Cost).
    \end{align}

We represent the present value of total cost in terms of $CONE$. Since $CONE$ is an annualized cost \footnote{Continuous discounting is a modeling convenience while summation denotes payments are settled discretely (eg, monthly) in practice.}, then 
\begin{align}
\label{eq:cost}
        PV(TotalCost) = \sum_0^{T+\tau}e^{-rt}CONE \cdot \Delta t.
\end{align}
This represents the present value of paying CONE continuously over [0,T] years, discounted at the rate r.

Under ROs as discussed in Section \ref{sec:assumptions}, the present value of the total revenue comes from:-
\begin{itemize}
    \item \textit{RO Premium (C)} : A one-time present-value payment (i.e. $\sum_{T}^{T+\tau} e^{-rt} \, \mathbb{E}\left[(S_t - K_t)^+\right] \, \Delta t$) at time 0.
    \item \textit{Energy revenues (R1) before the RO contract ($0 \text{ to } T$)}: Selling electricity freely at market prices $S_t$ amounts to $\sum_{0}^{T} e^{-rt} \, \mathbb{E}[S_t] \, \Delta t$.
    \item \textit{Energy revenues (R2) during the RO contract period ($T\text{ to }T+\tau$)}: Selling electricity at the capped strike price $K_t$ i.e. at $\min\{S_t, K_t\}$, aggregates to a value of $\sum_{T}^{T+\tau} e^{-rt} \, \mathbb{E}\left[\min(S_t, K_t)\right] \, \Delta t$.
\end{itemize}


Therefore, if we simplify $CONE$ into capital expenditure ($CapEx$) and $O\&M$, and break down the revenue into its constituent terms, 
and  apply the identity $\min(S_t, K_t) + (S_t - K_t)^+ = S_t$, then the Eq \ref{eq:breakeven} simplifies to:
\begin{align}
\label{eq:simple_break}
\sum_0^{T+\tau} e^{-rt} \mathbb{E}[S_t] \,\Delta t = \text{CapEx} + \text{O\&M} \cdot \sum_0^{T+\tau} e^{-rt} \Delta t
\end{align}

We, then use Eq. \ref{eq:simple_break} to calculate the minimum required contract duration. The general break-even condition above does not admit a closed-form solution for $\tau$. Therefore, given any expected revenue path, the present value of revenue and cost can be evaluated numerically, and the contract duration $\tau^*$ can be solved iteratively by increasing the value of $\tau$.

\section{Reliability Metrics and Strike Price}
\label{sec:metrics}
As power systems evolve, driven by high shares of renewable energy and decentralized generation, so too must the metrics used to evaluate capacity adequacy. This section introduces and motivates a shift from traditional adequacy metrics like Expected Energy Unserved (EEU), toward a more risk-aware, distribution-sensitive metric, namely Conditional Value-at-Risk (CVaR), to evaluate reliability and help in capacity procurement. Additionally, here, we discuss how risk aversion can be adjusted by altering the strike price, another key parameter set for calculating the premium.

Here, first we formally define unserved energy (energy defecit) or excess demand at any time $t$ under a capacity portfolio $\mathcal{R}$ as $Z_t(\mathcal{R}) = D_t - G_t(\mathcal{R})$, where $D_t$ is the total energy demand and $G_t$ includes the available renewable and conventional generation respectively. Let there be $n$ hours in the contract period $\tau$, for which we assess reliability.

\subsection{Conditional Value-at-Risk (CVaR)}
We begin by recalling the definition of CVaR \cite{mcneil2015quantitative}. Let $Z$ be a random variable that represents the total unserved energy in the evaluation period, i.e., $ Z = \sum_{t=1}^n [Z_t]^+$. We define $F_Z(.)$ as the cumulative distribution function (CDF) of $Z$. 

\begin{definition}[Conditional Value-at-Risk (CVaR)]
     Let $Z$ be a continuously distributed random variable with finite expectation i.e. $\mathbb{E}[|Z|] < \infty$ and let $\alpha \in (0,1)$. We define $CVaR_\alpha$ as the conditional expectation of $Z$ given that it exceeds the $\alpha-quantile$ :
    \begin{align*}
        \text{CVaR}_\alpha(Z) 
        &= \mathbb{E}[Z \mid Z \geq q_\alpha(Z)] 
        = \frac{1}{1 - \alpha} \int_\alpha^1 q_u(Z)\, du
    \end{align*}   
\end{definition}

Intuitively, $CVaR_\alpha$ is the EEU in the worst $1- \alpha$ fraction of the assessment period, where $EEU = \mathbb{E}\left[ \sum_{t=1}^n [Z_t]^+ \right]$.

\begin{proposition}[CVaR representation]
\label{cvarrepresentation} 
    As $\alpha \to 0$, $CVaR_\alpha \to EEU$ and as $\alpha \to 1$, $CVaR_\alpha \to \operatorname*{ess\,sup} Z$. This corresponds to the worst case (most extreme) blackout level.
\end{proposition}

\subsection{Reliability Options align with CVaR of Shortfall}
\begin{lemma}[RO premium as price–tail CVaR]
\label{lem:ROasCVAR}
Let $S_t$ be continuously distributed energy spot prices with finite expectation at each delivery time $t \in [T,T+\tau]$, and fix $\alpha \in (0,1)$. Choose the strike so that $K_t = q_\alpha(S_t)$. Then, the RO premium of Proposition~\ref{ROasEO} can be rewritten in the form of a linear combination of $\mathrm{CVaR}_\alpha(S_t)$ as:
\begin{align}
C(0,S_0)
    &= \sum_{t=T}^{T+\tau} e^{-rt}(1-\alpha)
       \bigl(\mathrm{CVaR}_\alpha(S_t) - K_t\bigr)\Delta t. \nonumber
\end{align}
\end{lemma}

\begin{definition}[Marginal value of lost load (VOLL)]
\label{def:vollmin}
Let $\varphi:\mathbb{R}_+\to\mathbb{R}_+$ be the scarcity map that converts
non-negative shortfall $z$ into a scarcity adder in price units (e.g., \$/MWh).
Then, we define the \emph{marginal VOLL} as $\mathrm{VOLL}\;\doteq\; \varphi'(z)$.
\end{definition}
Economically, $\varphi'(z)$ is the marginal scarcity price per additional unit of shortfall and  $\mathrm{VOLL}>0$ asserts that any positive shortfall raises the price by at least a fixed slope. 

\begin{theorem}[Alignment of RO premium with CVaR of shortfall]
\label{thm:RO_shortfall_CVaR}
Fix $\alpha \in (0,1)$ and a delivery time $t$. Under the scarcity-pricing Assumption~\ref{ass:scarcity} i.e., $S_t = s_{\text{base},t} + \varphi\bigl(Z_t^{+}(\mathcal{R})\bigr)$, if $  z_\alpha := q_\alpha\bigl(Z_t^{+}(\mathcal{R})\bigr)$, then 
$K_t := q_\alpha(S_t) = s_{\text{base},t} + \varphi(z_\alpha)$. Therefore, the RO premium in Proposition~\ref{ROasEO} can be written as, $C(0,S_0)$ =
\begin{align}
\label{eq:premium-shortfall-CVaR}
 \sum_{t=T}^{T+\tau} e^{-rt}VOLL(1-\alpha)
       \,
     \Bigl(\mathrm{CVaR}_\alpha\bigl(Z_t^{+}(\mathcal{R})\bigr) - z_\alpha\Bigr)\Delta t. 
\end{align}
\end{theorem}

\vspace{-8pt}

\section{Reliability Options as Capacity Auctions}
\label{sec:capacityauctions}
The contractual form of capacity auctions and reliability options on cleared resources diverges in several ways. 
\begin{definition}[CA Payment Stream] \label{def:CM}
Under a capacity auction, a generator receives the deterministic payment $\Pi^{\textrm{CA}}_{\text{cap}} = M\,Q,$
where $M$ is the per unit capacity payment in CA, and, when dispatched at time $t$, it earns the spot energy revenue of
\[
\Pi^{\textrm{CA}}_{\text{eng}}(t)=S_t\,q_t,  \qquad 0\le q_t\le Q.
\]
Total payoff in the delivery period of $\tau$ is : $\Pi^{\textrm{CA}}=\Pi^{\textrm{CA}}_{\text{cap}}+\sum_0^{\tau} e^{-rt} \Pi^{\textrm{CA}}_{\text{eng}}(t)\,\Delta t$.
\end{definition}

\begin{definition}[RO Payment Stream] \label{def:RO}
With an RO, the generator receives the upfront premium $\Pi^{\textrm{RO}}_{\text{prem}} = C\,Q,$
and energy payments as 
\[
\Pi^{\textrm{RO}}_{\text{eng}}(t)= q_t \min(S_t,K)\qquad 0\le q_t\le Q.
\]
Hence, total payoff $\Pi^{\textrm{RO}}=\Pi^{\textrm{RO}}_{\text{prem}}+\sum_0^{\tau} e^{-rt} \Pi^{\textrm{RO}}_{\text{eng}}(t)\,\Delta t$.
\end{definition}

\begin{proposition}[Consumers favour RO over CA]
\label{thm:dominance_capped}
Suppose ISO sets the RO premium at the CA payment, i.e.\ $C=M$.  Then,
\[
\mathbb{E}\!\bigl[\Pi^{\mathrm{RO}}\bigr]
\;<\;
\mathbb{E}\!\bigl[\Pi^{\mathrm{CA}}\bigr],
\qquad
\text{whenever } \Pr[S_t>K_t]>0.
\]

\end{proposition}
\begin{remark}[Overcompensation by CA]
\label{remark:overcompensate}
The proposition formalizes that \emph{capacity payments \& uncapped locational marginal prices} over-compensate suppliers relative to the consumer’s risk, whereas RO contracts right-size the transfer by internalizing scarcity-price risk.
\end{remark}

\vspace{-10pt}

\section{Modeling Prices as Stochastic Processes and  Premium as Option Derivative}
\label{sec:methodology}
In this section, we present the stochastic models, 
focusing on Markov Regime Switching Process (other models are mentioned in Appendix \ref{appdx:stoch}), to represent the underlying dynamics of wholesale electricity prices. These models are essential to capture key characteristics of electricity markets such as pronounced volatility, sudden structural breaks, and abrupt price spikes - that critically influence the valuation of capacity market premia. Price trajectories vary significantly across different electricity markets (Figure~\ref{fig:price_trends_all_singlecolumn}). It is evident that no single stochastic model, even with parameter adjustments, is sufficient to capture the full range of observed behaviors. Relying on a model calibrated to a calm historical period risks severely underestimating market risk - a limitation seen in much of the prior literature \cite{ANDREIS2020104705}, which often fits a single process to a narrow time window.


To address this, we adopt a Markov Regime-Switching framework. Here, a latent Markov chain captures regime shifts, while the price evolution within each regime is modeled using the most appropriate process. This flexible architecture allows the framework to adapt seamlessly between stable and stressed conditions, avoiding the pitfalls of one-size-fits-all modeling. 


\subsection{Distribution Comparison Metrics}
Here, we first discuss the statistical distance measures to compare simulated (\(S\)) vs empirical (\(P\)) wholesale energy market price distributions, where prices ($S_t$) are expressed in \$/MWh. The following statistical distances are used:

\begin{itemize}
    \item \textit{Weighted KL Divergence ($WeightedKL_{right}$):}
    \[
    D_{\text{KL}}^{(w)}(P \,\|\, S) = \sum_{x} w(x) \, P(x) \log \frac{P(x)}{S(x)}, 
    \]
    \[
     w(x) > 1 \;\; \text{if } x \in \text{tail}
    \]
    It is a tail sensitive extension of Kullback–Leibler (KL) divergence, where higher weights are assigned to the right tail of the distribution. 
    
    
    \item \textit{Tail Wasserstein (TailWass):}
    \[
    W_{\alpha}(P, S) = \inf_{\gamma \in \Gamma(P,S)} \int_{x \in \text{tail}_\alpha} |x - y| \, d\gamma(x,y)
    \]
    It is a modification of the Wasserstein distance restricted to the upper tail (e.g., prices above the 95th percentile). 
\end{itemize}

\subsection{Markov Regime Switching Model (MRSM)}
\label{sec:regimeswitchingmodelsection}
Here, we discuss the Markov Regime Switching Model to represent historical prices through distinct states or “regimes”, each governing a different phase of the price process, and assume that the future prices will evolve within one of these estimated regimes.

\begin{definition}[Markov–Switching $\operatorname{AR}(p)$ model]
\label{def:MS_AR_p}

Let $\{S_t\}_{t\ge 0}$ be the (observed) wholesale-price series and $\{R_t\}_{t\ge 0}$ an unobserved, finite‐state Markov chain taking values in the set of regimes $\{1,\dots,R\}$.
The process (of order $p$) is specified by the pair of equations:
\begin{align}
    S_t = \mu_{R_t} + \sum_{i=1}^{p}\phi_{i,R_t} S_{t-i}+\epsilon_t \label{eq:AR} \\
    \mathbb{P}(R_t = j| R_{t-1}=i) = \pi_{ij}, \quad \forall i,j \in \{1,\cdots, R\} \label{eq:MC}
\end{align}
    
with \( \mu_{r} \) as regime-specific intercept (long-run mean); \( \phi_{i,r} \)  is the autoregressive coefficients of order \(i=1,\dots,p\) in regime \(r\); \( \varepsilon_{t} \) represents i.i.d.\ innovations \( \varepsilon_{t}\sim\mathcal N\!\bigl(0,\sigma_{r}^{\prime\,2}\bigr) \) whenever \(R_t=r\); and \( \bm{\Pi}=[\pi_{ij}] \) is an $R\times R$ transition matrix of the latent Markov chain which captures the transition probability from one regime to the other.
Equation~\eqref{eq:AR} states that \emph{conditional} on the regime $R_t=r$ the prices follow an $\operatorname{AR}(p)$ with parameters $(\mu_{r},\phi_{1,r},\dots,\phi_{p,r},\sigma_{r}^{\prime\,2})$. Equation~\eqref{eq:MC} lets the model switch stochastically between regimes, thereby capturing structural breaks, volatility bursts, or changes in mean reversion that are typical of electricity markets.
\end{definition}

Now, even though we observe $S_t$, when we sign the contract (i.e., $t=0$), we do not observe the latent regime state $R_t$ at any time $t$ after signing the contract. So, our goal is to infer $R_t$ based on the historical energy prices. This follows the Hamilton Iterative Filtering Algorithm, which is described in Appendix \ref{sec:markov regime switching process}. To get an idea of the number of regimes existing in the historical dataset,  we use K-Means clustering and also infer the latent regime state $R_t$ for all times $t$. 

{\begin{algorithm}[!t]
\caption{Monte Carlo for Markov Regime-Switching OU}
\label{alg:MC_RegSw_OU}
\begin{algorithmic}[1]
\Require
  \Statex $T, \tau$: start and length of capacity delivery windo.w
  \Statex $r$: risk-free rate.
  \Statex $K$: strike (capacity cap price).
  \Statex $S_0, R_0$: initial value of the OU process and regime.
  \Statex $\{\kappa_r, \theta_r, \sigma_r\}$: OU parameters for each regime $r$.
  \Statex $\pi$: transition matrix or generator for Markov chain $\{R_t\}$
  \Statex $N_{\text{paths}}$: number of Monte Carlo sample paths
  \Statex $N$: number of time steps from $T$ to $T + \tau$

  \State $\Delta u_{\text{pre}} \gets \frac{T}{N_{\text{pre}}}$ \Comment{Time step from $0$ to $T$}
  \State $\Delta u \gets \frac{\tau}{N}$ \Comment{Time step from $T$ to $T+\tau$}
  \State \texttt{capacityPremium} $\gets 0$
  \For{$p = 1$ to $N_{\text{paths}}$}
    \State $(S, R) \gets (S_0, R_0)$
    
    \For{$k = 1$ to $N_{\text{pre}}$}
      \State $R \gets \text{sample\_next\_regime}(R, \pi, \Delta u_{\text{pre}})$
      \State $S \gets \text{OU\_step}(S, \kappa_R, \theta_R, \sigma_R, \Delta u_{\text{pre}})$
    \EndFor

    \State $\text{sumPayoff} \gets 0$
    \For{$k = 0$ to $N$}
      \State $u_k \gets T + k \cdot \Delta u$
      \State $R \gets \text{sample\_next\_regime}(R, \pi, \Delta u)$
      \State $S \gets \text{OU\_step}(S, \kappa_R, \theta_R, \sigma_R, \Delta u)$
      \State $\texttt{payoff} \gets \max(S - K,\, 0)$
      \State $\texttt{discf} \gets \exp(-r \cdot u_k)$
      \State $\text{sumPayoff} \gets \text{sumPayoff} + \texttt{payoff} \cdot \texttt{discf} \cdot \Delta u$
    \EndFor
    \State $\texttt{pathPremium}[p] \gets \text{sumPayoff}$
  \EndFor
  \State $\texttt{capacityPremium} \gets \frac{1}{N_{\text{paths}}} \sum_{p=1}^{N_{\text{paths}}} \texttt{pathPremium}[p]$
  \State \Return $\texttt{capacityPremium}$
\end{algorithmic}
\end{algorithm}
}

Next, for each regime $r \in R_t$, we couple the MRSM model with the mean-reverting OU process. Therefore, the final capacity premium charged at time $t=0$ is given by: $C(0,S_0) = \sum_{t=T}^{T+\tau} e^{-rt} \mathbb{E}[\{S_t - K\}^+ | S_0, R_0]\Delta t$. But now, $\mathbb{E}[\{S_t - K\}^+ | S_0, R_0]$ must account for all possible regime paths that the Markov chain $R_t$ may take between time 0 and t. Because each regime has different OU parameters, we can not have a closed-form solution for the model. The distribution of $S_t$ being a mixture of normals conditional on the path of regimes makes it all the more difficult. 

Therefore, we use Monte Carlo Simulation as described in Algorithm \ref{alg:MC_RegSw_OU} to simulate many paths of $\{R_t\}$ and $\{S_t\}$ from $t= T$ to $T+\tau$, to come up with the final capacity premium.

\section{Experiments and Results}
\label{sec:experiments}
\begin{figure}[!t]
    \centering
    

    \begin{subfigure}[b]{\linewidth}
        \centering
        \includegraphics[width=\linewidth]{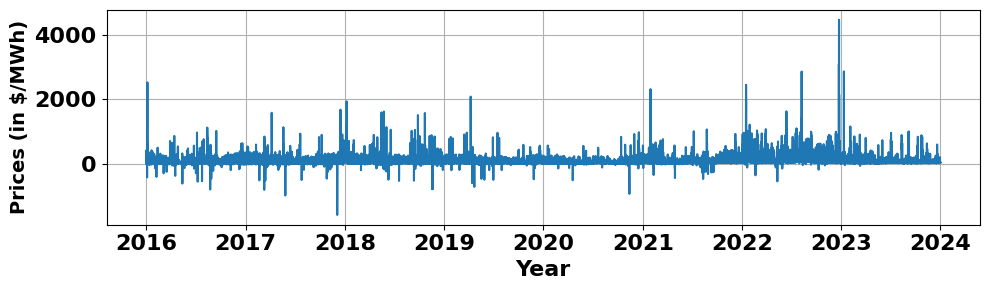}
        \caption{\small{New York (NYISO): Energy price series from 2016 to 2024. Sharp spikes, alongside brief negative prices are visible despite lower average volatility.}}
        \label{fig:prices_nyiso}
    \end{subfigure}
    
    \vspace{1em}

    



    \begin{subfigure}[d]{\linewidth}
        \centering
        \includegraphics[width=\linewidth]{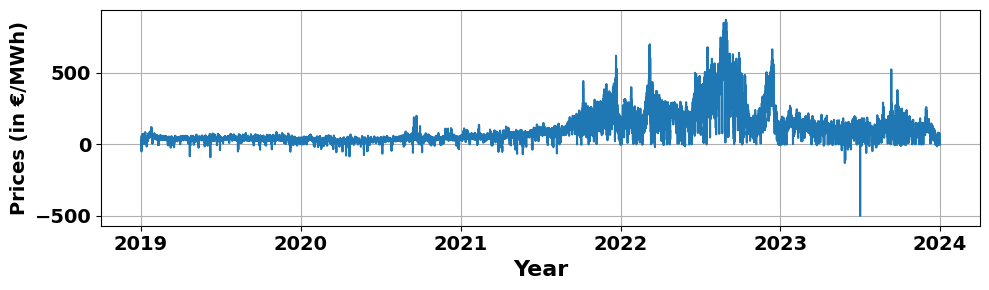}
        \caption{\small{Germany: Day-ahead prices from 2019 to 2024, capturing long-term stability and recent disruptions from 2022 onwards.}}
        \label{fig:prices_germany}
    \end{subfigure}
    


    \caption{Electricity prices across NYISO, and  Germany.}
    \label{fig:price_trends_all_singlecolumn}
    \vspace{-15pt}
\end{figure}

For our analysis, we retrieve publicly available electricity price (e.g., see Figure \ref{fig:price_trends_all_singlecolumn}), consumption, and generation data from multiple regional sources. 
(i) Data for New York is sourced from the NYISO  \href{https://www.nyiso.com/energy-market-operational-data}{Settlement archives}, which include historical 5-minute cleared real-time price information and capacity allocations monthly. These prices are averaged across 11 load zones, weighted by zonal loads at each hour. (ii) In the case of CAISO, we use hourly day-ahead price data from the \href{https://www.eia.gov/electricity/wholesalemarkets/data.php?rto=caiso}{U.S. Energy Information Administration (EIA)} and average them across three zones (NP-15, SP-15, and ZP-26). (iii) Similarly, we use the average day-ahead bus price data from the \href{https://www.eia.gov/electricity/wholesalemarkets/data.php?rto=ercot}{EIA} to model Electric Reliability Council of Texas (ERCOT) prices. (iv) German energy price data is retrieved from the \href{https://www.smard.de/home/downloadcenter/download-marktdaten/}{SMARD (Strommarktdaten) platform}, operated by the German Federal Network Agency. We use hourly day-ahead prices for the DE/LU bidding zone. (v) For Italy, we rely on day-ahead price and dispatch data from the \href{https://dati.terna.it/download-center#/fabbisogno/fabbisogno-italia}{Gestore Mercati Energetici} (GMI - the Italian Power Exchange). 
\vspace{-10pt}

\subsection{Case Study: Capacity Premium in Germany} 

The German spot market offers a vivid natural experiment in \emph{exogenous regime changes}: (i) the 2018-2021 oversupply of renewables that depressed prices to negative values and (ii) the price increase for 2021-2023. Each episode created a distinct state characterized by a dramatically different mean value and volatility that an ISO must price into forward capacity contracts. We use SMARD hourly prices from January 2019 to March 2024 (Figure \ref{fig:price_trends_all_singlecolumn}b). 


\paragraph{Rolling-window features and K-Means Clustering.}
To capture these structural breaks, we engineer a 30-day rolling \emph{Mean}, \emph{Volatility}, and \emph{Log-Return}. The elbow test on $k$-means inertia produces a value of $k=3$ as the best number of clusters observed from the data. Therefore, we next use Markov Autoregression using 3 regimes.

\begin{figure}[t]
    \centering
    \begin{subfigure}[t]{0.48\linewidth}
        \centering
        \includegraphics[width=\linewidth,height=1.5in]{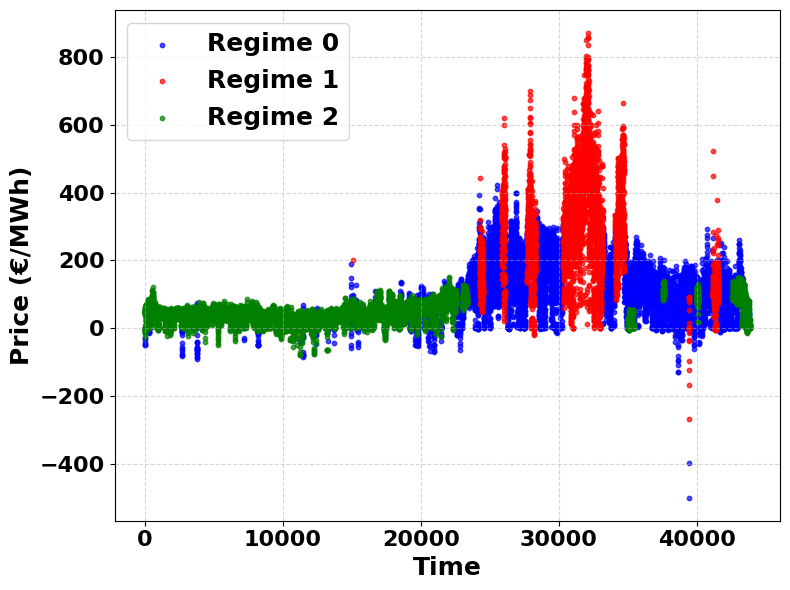}
        \caption{Regimes from MRSM.}
        \label{fig:regimes_MAR}
    \end{subfigure}
    \hfill
    \begin{subfigure}[t]{0.48\linewidth}
        \centering
        \includegraphics[width=\linewidth,height=1.5in]{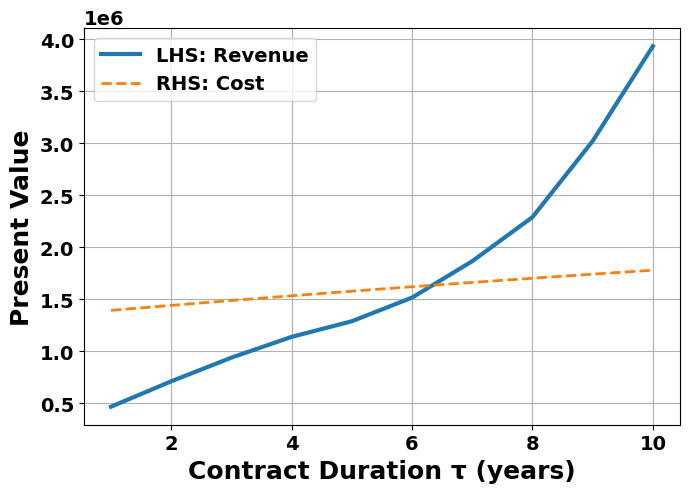}
        \caption{Contract Duration: $\tau \sim 6$ yrs.}
        \label{fig:mincontractdur}
    \end{subfigure}
    \caption{MRSM regimes and contract-duration with $CapEx = 1{,}290{,}806$~\text{€}/MW, and $O\&M = 52$k~\text{€}/MW-yr.}
    \vspace{-15pt}
\end{figure}

\paragraph{Markov Auto-Regession.} 
From the Bayesian Information Criterion (BIC) (we check this for each regime), we see a strong decay after lag 2 but significant dependencies till a lag of 24. Now, Markov Autoregression requires conditioning on both past observations and past regimes. The likelihood involves terms of the form: $P\left(S_t \middle|S_{t-1}, \dots, S_{t-p},R_t, \dots, R_{t-p}\right)$, which depend on an unobserved sequence of latent regimes. As a result, maximum-likelihood estimation (MLE) suffers from exponential path explosion: for lag order $p$ and $k$ regimes, the likelihood marginalizes over $k^{p+1}$ possible regime paths, rendering direct estimation computationally prohibitive.
So, we stick to a lower order of 2. We, therefore, restrict our attention to the AR($p$) model with $p=2$ and run Markov AutoRegression with 3 state regimes (Figure \ref{fig:regimes_MAR}). The parameters for each of the regimes are summarized in Table \ref{tab:parametersMarkov}.

\begin{table}[!t]
\begin{center}
\caption{Parameters for the 3-state AutoRegression Process.}
\label{tab:parametersMarkov}
\footnotesize
\resizebox{\linewidth}{!}{%
\begin{tabular}{| c | c  c  c  c | c  c |}
\hline
\textbf{Regime} & Constant & $\sigma'^2$ & AR(1) & AR(2) & Mean & Var \\
\hline
{\color{green}2 (Low)}  & 84.02  & 21.42   & 1.48 & -0.48 & 41.89  & 531.30 \\
{\color{blue}0 (Med)}   & 115.07 & 214.24  & 1.47 & -0.52 & 115.66 & 4622.64 \\
{\color{red}1 (High)}   & 326.55 & 1174.99 & 1.51 & -0.56 & 301.45 & 23725.24 \\
\hline
\end{tabular}}
\end{center}
\vspace{-10pt}
\end{table}

\begin{table}[!t]
\begin{center}
\caption{OU parameters, from AR-derived matrices.}
\label{tab:OUparams}
\begin{tabular}{| c | c | c | c |}
\hline
\textbf{Regime} & $\kappa$ & $\theta$ & $\sigma$ \\
\hline
{\color{green}2 (Low)} & 264.94 & 41.89  & 531.05 \\
{\color{blue}0 (Med)} & 324.25 & 115.66 & 1730.65 \\
{\color{red}1 (High)} & 320.45 & 301.45 & 3897.34 \\
\hline
\end{tabular}
\end{center}
\vspace{-15pt}
\end{table}

\paragraph{OU Process Estimation for each regime.} We then compute the MLE parameters of the underlying OU process for each regime, which is described in Table \ref{tab:OUparams}.


\paragraph{Calculating Contract Duration $\tau$.} From Section \ref{sec:contractduration}, we confirm that under a properly calibrated RO contract for a natural gas combined cycle (NGCC) based energy producer, $\tau =6$ years is necessary and sufficient for the cost recovery of $CapEx$ and $O\&M$ expenditures (Figure \ref{fig:mincontractdur}) \footnote{The CapEx anchor \$1.291M/MW (\$1,291/kW) sits within recent NGCC capital-cost ranges (\href{https://www.lazard.com/media/eijnqja3/lazards-lcoeplus-june-2025.pdf}{Lazard LCOE 2025, see p.~36}: \$1,200–\$1,600/kW). The $O\&M$ assumption \$52k/MW-yr is conservative but consistent with U.S. capacity-market studies when firm gas, property tax, and insurance are included (PJM 2025 CONE mentioned \href{https://www.pjm.com/-/media/DotCom/committees-groups/committees/mic/2025/20250221-special/pjm-qr-cone-and-vrr-curve-deck.pdf}{here} says levelized fixed O\&M $\approx$ \$38–\$60/kW-yr across zones).}.

\paragraph{Premia Charged.} The final capacity premium charged at time $t=0$, for T = 1 yrs, $\tau$ = 6 yrs, K = 150 \text{ €}/MWh, r = 2.64\%, is then given by:
\begin{align*}
\underbrace{C(0,S_0)}_{\frac{\text{€}}{MW}}  &= \sum_{t=T}^{T+\tau} e^{-rt} \underbrace{\mathbb{E}[\{S_t - K\}^+ | S_0, R_0]}_{\frac{\text{€}}{MWh}} \underbrace{\Delta t}_{1 hr}\\
&= 294,775.92 \text{ €}/MW
\end{align*}
Let's denote $A$ as the levelized annual premium (€/MWyr). Because the premium is paid at the \emph{start} of each year, its present value is $C =\sum_{n=0}^{\tau-1} A\,e^{-r n}$. Hence the levelized annual payment is $A=C/({\sum_{n=0}^{\tau-1} e^{-r n}})$. If instead the premium were paid \emph{continuously} over the interval \([T,T+\tau]\), the equivalence condition becomes
$
A =\;Cr/({1-e^{-r\tau}})
$. This yields the premium as $\text{€ } 52,428 /MWyr$.
\begin{figure}
    \centering
    \begin{minipage}[t]{0.23\textwidth}
        \centering
        \includegraphics[width=1\textwidth]{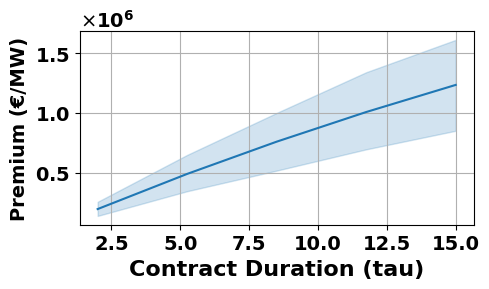}
    \end{minipage}%
    \begin{minipage}[t]{0.23\textwidth}
        \centering	\includegraphics[width=1\textwidth]{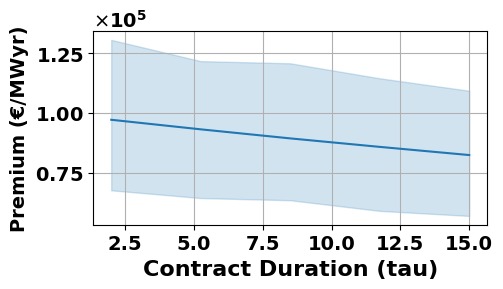}
    \end{minipage}

    \caption[Sensitivity Analysis]{Sensitivity Analysis of Premium with respect to $\tau$, by varying $T$, $r$ and $K$. \label{fig:senstivity_tau}}
    \vspace{-10pt}
\end{figure}

\begin{figure}[!t]
    \centering
    \begin{minipage}[t]{0.23\textwidth}
        \centering
        \includegraphics[width=1\textwidth]{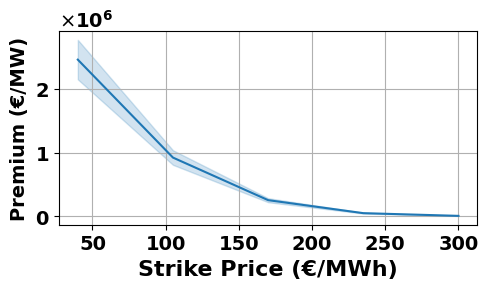}
    \end{minipage}%
    \begin{minipage}[t]{0.23\textwidth}
        \centering	\includegraphics[width=1\textwidth]{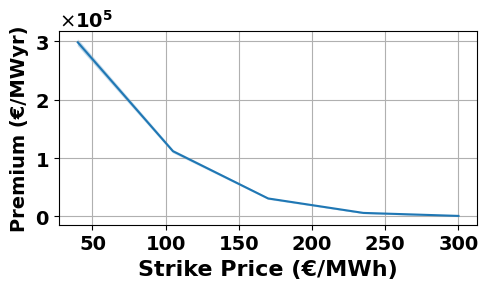}
    \end{minipage}

    \caption[Sensitivity Analysis]{Sensitivity Analysis of Premium with respect to $K$, by varying $T$, $r$ and $\tau$. \label{fig:senstivity_K}}
    \vspace{-10pt}
\end{figure}
\paragraph{Sensitivity Analysis of Capacity Premium in Germany.}
To understand how market design parameters affect the reliability option premium in Germany, we conduct a sensitivity analysis with respect to contract duration ($\tau$) and strike price ($K$). Figure \ref{fig:senstivity_tau} and \ref{fig:senstivity_K} show that the premium is highly sensitive to both factors. Longer contract durations yield significantly higher premia, as they ensure more prolonged protection against scarcity events, increasing the expected payoff of the option. However premium charged per year for each MW decreases with the increase in contract duration, which proves the sub-linear growth of premium values. 

Higher strikes make the option cheaper since it triggers less often and pays less when it does, especially in Germany, where outsized spikes are rare. The shaded area shows how results change when we vary contract length, discount rate, and the threshold. On the right, we show the same relationship but annualized to € per MW-year; annualizing removes most of the variation.
These relationships are nonlinear and emphasize the need for regulators to carefully calibrate $\tau$ and $K$ to ensure investor incentives while guaranteeing system adequacy. 

\subsection{Empirical Benchmarking - Comparison of ROs with Existing Mechanisms in different electricity markets}

\begin{table}
\begin{center}
\caption{Goodness-of-fit metrics comparing simulated price distributions against observed prices across regions.}
\label{tab:distmetrics}
\footnotesize
\begin{tabular}{| c | c | c | c |}
\hline
\textbf{Region} & \textbf{Model Name} & \textbf{$WeightedKL_{right}$} & \textbf{TailWass} \\
\hline

\multirow{3}{*}{\shortstack{Texas \\ (ERCOT)}} 
& MRSM-OU      & 1.73      & $\times$ \\
& OU-jumps     & 0.09      & 454.12 \\
& \textbf{GARCH} & \textbf{0.05} & \textbf{147.93} \\
\hline

\multirow{3}{*}{\shortstack{New York \\ (NYISO)}} 
& MRSM-OU      & 0.82      & $\times$ \\
& OU-jumps     & 0.0014    & 51.04 \\
& \textbf{GARCH} & \textbf{0.0009} & \textbf{6.44} \\
\hline

\multirow{3}{*}{\shortstack{California \\ (CAISO)}} 
& MRSM-OU      & 2.29      & $\times$ \\
& \textbf{OU-jumps} & \textbf{0.013} & \textbf{48.19} \\
& GARCH        & 0.017     & 79.27 \\
\hline

\multirow{3}{*}{Italy} 
& \textbf{MRSM-OU} & \textbf{0.008} & \textbf{9.71} \\
& OU-jumps     & 0.063     & 95.45 \\
& GARCH        & 0.041     & 78.28 \\
\hline

\multirow{3}{*}{Germany} 
& \textbf{MRSM-OU} & \textbf{0.21} & \textbf{46.21} \\
& OU-jumps     & 0.26      & 79.84 \\
& GARCH        & 0.32      & 318.33 \\
\hline

\end{tabular}
\end{center}
\vspace{-20pt}
\end{table}

\begin{table*}[htbp]
\begin{center}
\caption{Comparing capacity pricing mechanisms across selected electricity markets. 
CRM payments for ROs reflect only premium values computed at the minimum contract duration for viability.}
\label{tab:crm_fullborder}
\footnotesize
\resizebox{\linewidth}{!}{%
\begin{tabular}{| c | c | c | c | c | c | c | c | c | c |}
\hline
\multirow{2}{*}{\textbf{Region}} 
& \multirow{2}{*}{\shortstack{\textbf{Model Fitted} \\  (with jumps)}} 
& \multirow{2}{*}{\textbf{Regime}} 
& \multicolumn{2}{c|}{\textbf{Spot Price Dynamics}} 
& \textbf{CRM} 
& \textbf{Strike Price (K)} 
& \textbf{Min. Contract} 
& \textbf{CRM Payment} 
& \textbf{Energy-only Rev} \\
\cline{4-5}
& & & $\mu$ & $\sigma$ & \textbf{type} & \textbf{(-/MWh)} & \textbf{Duration ($\tau$ yrs)} & \textbf{(-/MWyr)} & \textbf{(-/MWyr)} \\
\hline

\multirow{2}{*}{\shortstack{Texas \\ (ERCOT)}} 
& \multirow{2}{*}{GARCH} & \multirow{2}{*}{--} 
& \multirow{2}{*}{56.02} & \multirow{2}{*}{380.45} 
& -- & -- & -- & -- & 417.5k \$ \\
& & & & & RO & 284.29 \$ & 5 & 71k \$ & 296k \$ \\
\hline

\multirow{2}{*}{\shortstack{New York \\ (NYISO)}} 
& \multirow{2}{*}{GARCH} & \multirow{2}{*}{--} 
& \multirow{2}{*}{36.44} & \multirow{2}{*}{49.90} 
& CA & -- & 8 & 23k \$ & 311k \$ \\
& & & & & RO & 90.91 \$ & 8 & 16k \$ & 297k \$ \\
\hline

\multirow{2}{*}{\shortstack{California \\ (CAISO)}} 
& \multirow{2}{*}{OU} & \multirow{2}{*}{--} 
& \multirow{2}{*}{51.13} & \multirow{2}{*}{57.74} 
& CO & -- & 1 & 100k \$ & 497k \$ \\
& & & & & RO & 135.71 \$ & 3 & 21k \$ & 450k \$ \\
\hline

\multirow{2}{*}{Italy} 
& \multirow{2}{*}{MRSM-OU} 
& R1 & 66.96 & 32.29 & RO & 85 € & 15 & 70k € & 728k € \\
& & R2 & 306.52 & 115.76 & RO & 85 € & 2 & 11k € & 728k € \\
\hline

\multirow{3}{*}{Germany} 
& \multirow{3}{*}{MRSM-OU} 
& R1 & 33.76 & 9.23 
& \multirow{3}{*}{\shortstack{SR \\ RO}} 
& \multirow{3}{*}{\shortstack{-- \\ 125 €}} 
& \multirow{3}{*}{\shortstack{2 \\ 6}} 
& \multirow{3}{*}{\shortstack{63k € \\ 52k €}} 
& \multirow{3}{*}{\shortstack{-- \\ 540k €}} \\
& & R2 & 82.84 & 27.86 & & & & & \\
& & R3 & 216.73 & 107.49 & & & & & \\
\hline

\end{tabular}}
\end{center}
\vspace{-20pt}
\end{table*}

Based on the tail-sensitive metrics from Table \ref{tab:distmetrics}, we select the best model for each region to compute the capacity premia.

Table~\ref{tab:crm_fullborder} compares our reliability-option (RO) valuations against two polar cases. 
At one end is the \emph{energy-only} paradigm - Texas (ERCOT), where generators rely solely on spot-market earnings. Given the distribution of Texas energy prices, a GARCH model with jumps provides the best fit. We set $\alpha=0.95$ to model the strike price, yielding $K = 284.29\$/\text{MWh}$. Break-even analysis indicates a minimum contract duration of five years for project viability through ROs. 
In theory, if the strike price $K$ and premium $C$ satisfy the “option-parity” identity used in Eq \ref{eq:simple_break}, the present value of an energy-only stream should equal that of an RO stream (premium plus capped energy). In practice, if we consider revenues earned across all operating hours for a five-year horizon, the levelized energy-only income rises to \$417.5k/MW-yr\footnote{If reserves fall dangerously low, ERCOT declares emergency conditions and deploys pre-arranged interruptible load resources, which are paid (check \href{https://www.ercot.com/services/programs/load/eils}{Emergency Response Services} for reference) to stop using power immediately. In our baseline calculations, we abstract out such emergency intervention payments. Explicitly modeling such payments would further narrow the gap between the baseline and ROs.}, with a present value of \$1.99M/MW, that is about 12\% higher than the \$1.75M/MW (combined capacity and energy revenues) obtained under ROs. This gap of 12\% reflects the model-driven error of the fitted stochastic process's representation of  ERCOT prices.
\\

At the other extreme are the existing capacity mechanisms:

(i) The New York Independent System Operator (NYISO) procures capacity through a statewide Installed Capacity (ICAP) market\footnote{It serves as the reference capacity zone for the entire state, but excluding transmission-constrained areas such as NYC and Zones H, I, and K. These constrained zones clear at higher prices due to local deliverability limits, while the statewide reference zone reflects unconstrained conditions.}. 
Based on NYISO historical ICAP data from 2011–2022, the average clearing price in the statewide reference zone is approximately \$23k/MW-yr, which we adopt as the capacity payment in our net present value (NPV) analysis \cite{nyiso_capacity_market}. From the break-even analysis, we calculate an eight-year contract duration (considering NYISO’s annual capacity obligation reset) for both ROs and CAs. Additionally, for ROs, we set a strike price of \$90.91/MWh (i.e., from $\alpha=0.95$). When compared to the RO premium estimate of \$16 k/MW-yr, the two mechanisms exhibit near-parity (the \$7k/MWyr gap in the CRM payments can be easily closed using an auction process). This suggests that, at the statewide level, an RO design could deliver comparable cost recovery for generators as the existing ICAP construct. The slightly higher energy revenue for CAs is a result of Remark~\ref{remark:overcompensate} indicating over-compensation.

(ii) California’s Capacity Obligation (CO) mechanism is implemented through the Resource Adequacy (RA) program, overseen by the California Public Utilities Commission \cite{CAISO2024_Section43A}. We use the 2022 Average Price\footnote{Check Section 4, Table 6 of the 2022 Resource Adequacy  Report \href{https://www.cpuc.ca.gov/-/media/cpuc-website/divisions/energy-division/documents/resource-adequacy-homepage/2022-ra-report_05022024.pdf}{here}.} of \$8.31/kW-month - assuming this rate is levelized across the year. Converting to a \$/MW-yr basis, we get \$ 99,720/MW-yr, which is much higher (almost $5\times$, unlikely that such a gap could be closed through auction adjustments alone) than the premium payment of \$21k /MW-yr estimated by ROs. We assume a one-year contract duration, as prices are recalibrated annually based on Load Serving Entity (LSE) filings, for COs. On the other hand, the strike price and the contract duration are set as \$ 135.71/MWh (from $\alpha=0.95$) and 3 years (from break-even analysis), respectively. Thus, this fixed capacity payment still leads to over-compensation, similar to our findings in the NYISO case. 

(iii) The Italian capacity market uses ROs as CRMs. They clear prices for existing (1-year contracts) and new capacity (15-year contracts) separately \cite{RSE_CapacityMarket_2024}. From the 2024 auction, a new capacity entrant receives €70,000/MW-yr. The scheme follows a reliability options design with a daily updated strike price linked to MGP-GAS and MI-GAS market prices; for simplicity, we hold it fixed at the peaker reference level of €85/MWh (2018–2019)\footnote{Check strike price on page 2 of the \href{https://www.arera.it/fileadmin/allegati/audizioni/pubbliche/19/EFET.pdf}{ARERA Strategic Plan 2019-2021}.}. Notably, this legacy RO tariff no longer reflects post-2022 volatility, potentially inflating premia. A regime-switching approach, thus, could better capture such structural shifts (although sufficient data per regime remains a limitation) and can bring down the cost to the consumers.

(iv) The German capacity framework operates under a strategic reserve (SR) model, structured into three reserve categories: grid reserves (primarily for redispatch), capacity reserves (for resource adequacy), and security standbys\footnote{Check \href{https://systemmarkt.net/Mediathek/20221207_Strategic_Reserves_SysteMarket.pdf}{here} for details on capacity remuneration in Germany.}. Capacity reserves\footnote{Check \href{https://www.bundesnetzagentur.de/DE/Fachthemen/ElektrizitaetundGas/Versorgungssicherheit/KapRes/start.html}{here} for details on capacity reserves.} are auction-based and are contracted for two-year periods. For the 2022–2024 performance period, the capacity reserve auction cleared at €62,940/MW-year\footnote{ Check \href{https://www.netztransparenz.de/en/Ancillary-Services/System-operations/Capacity-reserve/Publications-on-the-2022-2024-performance-period}{here} for the levy value for capacity reserve plants.}. Participation in the SR mechanism requires generators to withdraw from the regular energy market, thus resulting in no revenues earned from the energy-only market. Comparing this with ROs, by setting a strike price of €125/MWh\footnote{See \href{https://www.arera.it/fileadmin/allegati/docs/17/592-17.pdf}{here} for details in latest economic parameters.} and contract duration of 6 years, we are able to achieve a higher net income (of almost €1.5M/MW) for the generators sufficient for cost recovery.
\\
Together, these findings highlight how well-calibrated ROs occupy a middle ground: they close the missing-money gap without imposing the systematic over-payments.

\section{Conclusion}
\label{sec:conclusion}

In summary, \capoptix\ is a financial options–based framework that calibrates reliability options to deliver realistic, risk-adjusted incentives. Using a modular stochastic engine built on regime-switching, Ornstein–Uhlenbeck, and jump–diffusion dynamics, we show through multi-region simulations for New York, California, Texas, Italy, and Germany that reliability options priced via \capoptix\ can recover costs while aligning generator incentives with system reliability, outperforming traditional mechanisms such as capacity auctions, strategic reserves, and centralized obligations in terms of preserving market signals and providing effective hedging against scarcity-driven price volatility. Our results indicate that \capoptix\ yields economically meaningful premia and contract durations that mitigate overpayment yet secure investment, offering regulators and system operators a transparent design tool.

\appendices

\section{Other Stochastic Models}
\label{appdx:stoch}
\subsection{Mean-Reverting Ornstein-Uhlenbeck (OU) Process}
\label{sec:mean-reverting}
In many electricity markets, especially those with relatively stable supply-demand dynamics, prices tend to exhibit mean-reverting behavior. 
We model it using an OU process  \cite{uhlenbeck1930theory}:
\begin{align}
    dS_t = \kappa(\theta - S_t)dt + \sigma dW_t, \forall t \in [0, T+\tau]
\end{align}
where, $\theta$ is the long-term mean and $\kappa > 0$ is the speed of mean reversion. We estimate these parameters by maximizing the log-likelihood of the observed data. 


\begin{lemma}[OU Process is Gaussian and Stationary]
\label{OULemma_gaussian}
The solution to the Ornstein–Uhlenbeck process is given by:
\begin{align}
\label{eq:OUprocessSt}
    S_t = S_0 e^{-\kappa t} + \theta(1 - e^{-\kappa t}) + \sigma \int_0^t e^{-\kappa(t - s)} dW_s. 
\end{align}
Then, \( S_t \sim \mathcal{N}(\mu_t, \sigma_t^2) \), where $\mu_t = S_0 e^{-\kappa t} + \theta(1 - e^{-\kappa t}),$ and $\sigma_t^2 = \frac{\sigma^2}{2\kappa}(1 - e^{-2\kappa t})$.
\end{lemma}



To model the infrequent energy price spikes, we now extend the OU model by introducing a jump component as follows:

\begin{assumption} [Jumps]
    Jumps have two components  -  

(i) \textbf{Size of jumps:} We assume the size of the jumps $Y_t$ are normally distributed (i.e. $ Y_t \sim \mathcal{N}(\mu_y, \sigma_y^2)$).

(ii) \textbf{Arrivals:} We assume that Jump arrivals ($N_t$) follow  $\text{Poisson} (\lambda)$ to count the number of jumps up to time \( t \), where \( \lambda \) denotes the average jump intensity. 
Therefore, the change in the number of jumps over a small interval is :
\[
dN_t =
\begin{cases}
    1, & \text{with probability  } (\lambda dt)\\
    0, & \text{with probability  } (1 - \lambda dt) 
\end{cases}
\]
The Jump component is independent to the Brownian Motion and the Jump size is independent of the arrivals.
\end{assumption}

This accounts for rare but significant price disruptions due to market shocks. Hence, 
\begin{align*}
    dS_t = \kappa(\theta - S_t) dt + \sigma dW_t +  Y_t \cdot dN_t.
\end{align*}

Due to the compound nature of the jumps and the absence of closed-form expressions for the option value, we estimate the capacity premium using Monte Carlo simulations.

\subsection{Markov Regime Switching Process}
\label{sec:markov regime switching process}
\begin{definition}[Filtered regime probability]
For each regime $j\in\{1,\dots,R\}$ define
\[
\varepsilon_{j,t} = \mathbb{P}(R_t = j | \Omega_t; \theta)  \qquad
   \Omega_t:=\{S_t,S_{t-1},\dots\},
\]
i.e.\ the posterior probability that the system is in regime $j$ after observing prices up to~$t$.
\end{definition}

In the above equation, $\theta$ represents the parameter set from historical price data. Each regime j is defined by the parameter set $\theta_j = \{\mu_j, \phi_{1,j},\cdots,\phi_{p,j}, \sigma_j^2 \}$. The full model includes - regime switching parameters $\{\theta_1, \cdots, \theta_R\}$ and transition probabilities $\pi_{i.j}$ as parameters. To estimate the parameters ($\theta$) of the model, we maximize the log-likelihood of the observed data. Since no closed-form solution exists, we use gradient-based methods like $L-BFGS$ to solve it.

\begin{lemma}[Hamilton filter recursion \cite{hamilton1989}]
\label{lem:hamilton}
Given filtered probabilities $\varepsilon_{i,t-1}$ for all $i$, the one-step prediction $\tilde\varepsilon_{j,t}$, observation likelihood $\mathcal L_t^{(j)}$, normalizing constant $L_t$, and posterior update $\varepsilon_{j,t}$ are:
\begin{align*}
\tilde\varepsilon_{j,t}
   &:=\sum_{i=1}^{R}\pi_{ij}\,\varepsilon_{i,t-1}, \quad \mathcal L_t^{(j)}
   :=\frac{1}{\sqrt{2\pi\sigma_j^{\prime 2}}}\,
     \exp\!\Bigl\{-\tfrac{(S_t-\hat S_t^{(j)})^{2}}{2\sigma_j^{\prime2}}
          \Bigr\},\\
L_t &:=\sum_{j=1}^{R}\tilde\varepsilon_{j,t}\,\mathcal L_t^{(j)}, \quad \varepsilon_{j,t}:=\frac{\tilde\varepsilon_{j,t}\,\mathcal L_t^{(j)}}{L_t},
\end{align*}
where
\(
   \hat S_t^{(j)}
   =\mu_j+\sum_{k=1}^{p}\phi_{k,j}S_{t-k}.
\)
\end{lemma}



    




\label{sec:oneorderMSARtoOU}
\begin{lemma}[Discrete--to--continuous mapping, single–order case]
\label{lem:AR1_to_OU}
Fix a regime $r\in\{1,\dots,R\}$ and sampling step $\Delta t>0$. Assume the price dynamics in that regime are observed at integer multiples of $\Delta t$ and satisfy the AR(1)
\[
   S_{t+\Delta t}
   \;=\;
   \mu_{r}
   +\phi_{1,r}\,S_{t}
   +\varepsilon_{t+\Delta t},
   \qquad
   \varepsilon_{t+\Delta t}\sim\mathcal N(0,\sigma_r^{\prime\,2}).
\]
Then there exists a scalar OU process for $s\in[t,t+\Delta t)$,  $dS_s\;=\;\kappa_r\!\bigl(\theta_r-S_s\bigr)\,ds +\sigma_r\,dW_s$,
with parameters (in terms of AR(1) parameters) given by :
\label{eq:OU_mapping}
\begin{align*}
     \kappa_r   &=-\frac{\ln\phi_{1,r}}{\Delta t}, &
     \theta_r   &=\frac{\mu_r}{1-\phi_{1,r}}, &
     \sigma_r   &=\frac{\sigma_r'}{\sqrt{\Delta t}}\,
                  \sqrt{\frac{2\ln\phi_{1,r}}{\phi_{1,r}^{2}-1}}.
\end{align*}
\end{lemma}

\section{Deferred Proofs}
\subsection{Proof of Proposition \ref{ROasEO}}
\begin{proof}
The capacity contract is a sequence (or strip) of European call options, each maturing at time \( t \in [T, T+\tau] \), with the same strike price \( K_t \), reflecting the capped energy price the LSE pays during shortage periods.

Each such option yields a payoff of \( (S_t - K_t)^+ \), exercised only when $S_t > K_t$. By the principle of risk-neutral valuation, the present value of a single option expiring at time \( t \) is:
\[
C_t = e^{-rt} \, \mathbb{E}^{\mathbb{Q}} \left[ (S_t - K)^+ \right]
\]

Since delivery happens across multiple periods \( t \in [T, T+\tau] \), the total capacity premium is the sum of all such option values, scaled by  \( Q \), yielding:
\[
C(0, S_0) = Q \sum_{t = T}^{T+\tau} C_t \Delta t = Q \sum_{t = T}^{T + \tau} e^{-rt} \, \mathbb{E}^{\mathbb{Q}} \left[ (S_t - K)^+ \right] \Delta t
\]
\end{proof}

\subsection{Proof of Proposition \ref{premium_as_netcone}}
\begin{proof}
    By Eq \ref{eq:breakeven}, Eq \ref{eq:cost}, and from Def \ref{def:netcone} we have :
            $C = \int_0^{T+\tau}e^{-rt}\text{CONE }dt - R1 - R2 \approx NetCONE$.
\end{proof}

\subsection{Proof of Proposition \ref{cvarrepresentation}}
\begin{proof}
     As $\alpha \to 0$, $q_0(Z) = 0$. Therefore,  $\text{CVaR}_0(Z) = \mathbb{E}[Z \mid Z \geq 0] = \mathbb{E}[Z] = EEU$. 
     
     Now, for $\alpha \to 1$, let $M^\ast := \operatorname*{ess\,sup} Z$. To prove the upper bound, fix $\varepsilon>0$. By definition of $M^\ast$ we have $\mathbb{P}(Z > M^\ast + \varepsilon)=0$, so $q_u(Z)\le M^\ast+\varepsilon$ $\forall u\in(0,1)$. Hence,
\[
\mathrm{CVaR}_\alpha(Z) = \frac{1}{1-\alpha}\int_\alpha^1 q_u(Z)\,du \le M^\ast+\varepsilon
\]
$\forall \alpha$, which yields $\limsup_{\alpha\uparrow 1}\mathrm{CVaR}_\alpha(Z)\le M^\ast$ as $\varepsilon\downarrow 0$.

To prove the lower bound, fix $\varepsilon>0$. Since $M^\ast$ is the smallest almost-sure upper bound, $M^\ast-\varepsilon$ is not an upper bound and $\mathbb{P}(Z > M^\ast-\varepsilon)>0$, so $u_0:=F_Z(M^\ast-\varepsilon)<1$.
For any $u>u_0$ one must have $q_u(Z)\ge M^\ast-\varepsilon$, and therefore, for all $\alpha>u_0$,
\[
\mathrm{CVaR}_\alpha(Z) = \frac{1}{1-\alpha}\int_\alpha^1 q_u(Z)\,du \ge M^\ast-\varepsilon.
\]
Thus $\liminf_{\alpha\uparrow 1}\mathrm{CVaR}_\alpha(Z)\ge M^\ast$ as $\varepsilon\downarrow 0$.
\end{proof}

\subsection{Proof of Proposition \ref{thm:dominance_capped}}
\begin{proof}
With $C=M$,
\begin{align*}
    \mathbb{E}\!\bigl[\Pi^{\mathrm{RO}}\bigr]
      &= M Q
        + \int_0^{\tau} e^{-rt}\,
            \mathbb{E}\!\bigl[\min(S_t,K)\,q_t\bigr]\,dt,\\
    \mathbb{E}\!\bigl[\Pi^{\mathrm{CA}}\bigr]
      &= M Q
        + \int_0^{\tau} e^{-rt}\,
            \mathbb{E}[S_t q_t]\,dt.            
\end{align*}

Because $\min(S_t,K)\le S_t$ pointwise and the inequality is strict with
positive probability when $S_t>K$, the integral part of the RO payoff is
strictly smaller, proving the result.
\end{proof}

\subsection{Proof of Lemma \ref{lem:ROasCVAR}}
\begin{proof}
     Given $S_t$, with CDF $F_{S_t}(s)$ and PDF $f_{S_t}(s)$ :
\begin{align*}
    \mathbb{E}[(S_t - K_t)^+] 
    &= \int (S_t -K_t) d\mathbb{P}(S_t > K_t)\\
    &= \int_{K_{t}}^\infty (s - K_t) f_{S_t}(s)ds \\
    &= \int_{K_{t}}^\infty sf_{S_t}(s)ds - \int_{K_{t}}^\infty  K f_{S_t}(s)ds\\
    &= (1-\alpha)CVaR_\alpha(S_t) - K_t (1- F_{S_t}(K_t))
\end{align*}
The last line is by the definition of $CVaR$, where $q_\alpha (S_t) = K_t$. Thus, by quantile definition, we have $F_{S_t}(K_t) = \alpha$. 
\end{proof}

\subsection{Proof of Theorem~\ref{thm:RO_shortfall_CVaR}}
\begin{proof}[Proof]
By the scarcity-pricing Assumption~\ref{ass:scarcity}, we have
$S_t = s_{\text{base},t} + \varphi\bigl(Z_t^{+}(\mathcal{R})\bigr)$.
On the relevant tail region, we assume a linear scarcity response with slope $\mathrm{VOLL}>0$, i.e., $ \varphi(z) = \varphi(0) + \mathrm{VOLL}\,z$, so that
\[
S_t = s_{\text{base},t} + \varphi(0) + \mathrm{VOLL}\,Z_t^{+}(\mathcal{R}).
\]

Since $S_t$ is an affine, strictly increasing function of
$Z_t^{+}(\mathcal{R})$, quantiles and CVaR transform linearly. Writing
$z_\alpha := q_\alpha(Z_t^{+}(\mathcal{R}))$, we obtain
\[
K_t = q_\alpha(S_t)
    = s_{\text{base},t} + \varphi(0) + \mathrm{VOLL}\,z_\alpha,
\]
\[
\mathrm{CVaR}_\alpha(S_t)
    = s_{\text{base},t} + \varphi(0)
      + \mathrm{VOLL}\,\mathrm{CVaR}_\alpha\bigl(Z_t^{+}(\mathcal{R})\bigr).
\]
Subtracting the two,
\[
\mathrm{CVaR}_\alpha(S_t) - K_t
    = \mathrm{VOLL}\,
      \Bigl(\mathrm{CVaR}_\alpha\bigl(Z_t^{+}(\mathcal{R})\bigr) - z_\alpha\Bigr).
\]

By Lemma~\ref{lem:ROasCVAR}, for this choice of $K_t$ we have
\begin{align*}
\mathbb{E}\bigl[(S_t-K_t)^+\bigr]
  &= (1-\alpha)\,\bigl(\mathrm{CVaR}_\alpha(S_t) - K_t\bigr)\\
  &= (1-\alpha)\,\mathrm{VOLL}\,
    \Bigl(\mathrm{CVaR}_\alpha\bigl(Z_t^{+}(\mathcal{R})\bigr) - z_\alpha\Bigr).
\end{align*}

Substituting this expression into the RO premium formula of Proposition~\ref{ROasEO} yields the expression of Theorem~\ref{thm:RO_shortfall_CVaR}. 
\end{proof}

\subsection{Proof of Lemma \ref{OULemma_gaussian}}
\begin{proof}
Let integrating factor be \( e^{\kappa t} \). Multiply both sides of the SDE by \( e^{\kappa t} \), and  apply product rule in stochastic calculus:
\[
d(e^{\kappa t} S_t) = e^{\kappa t} dS_t + \kappa e^{\kappa t} S_t dt = e^{\kappa t} \kappa \theta\, dt + e^{\kappa t} \sigma\, dW_t.
\]

Therefore, we integrate both sides from \( 0 \) to \( t \) to get $S_t = \mu_t + \sigma \int_0^t e^{-\kappa(t - s)} dW_s$, where \( \mu_t = S_0 e^{-\kappa t} + \theta(1 - e^{-\kappa t}) \).





The stochastic integral \( \int_0^t e^{-\kappa(t - s)} dW_s \) is a normally distributed random variable with mean zero and variance:
\[
\mathrm{Var}\left(\int_0^t e^{-\kappa(t - s)} dW_s\right) = \int_0^t e^{-2\kappa(t - s)} ds = \frac{1 - e^{-2\kappa t}}{2\kappa}.
\]

Therefore, the variance of \( S_t \) is: $\sigma_t^2 = \sigma^2 \cdot \frac{1 - e^{-2\kappa t}}{2\kappa}$.

Hence, we conclude $S_t \sim \mathcal{N}(\mu_t, \sigma_t^2)$, where $\mu_t = S_0 e^{-\kappa t} + \theta(1 - e^{-\kappa t})$, and $\sigma_t^2 = \frac{\sigma^2}{2\kappa}(1 - e^{-2\kappa t})$.

As \( t \to \infty \), \( S_t \to \mathcal{N}(\theta, \frac{\sigma^2}{2\kappa}) \), so the process is asymptotically stationary with limiting mean \( \theta \) and variance \( \frac{\sigma^2}{2\kappa} \).
\end{proof}

\bibliographystyle{IEEEtran}
\bibliography{references} 

@article{hamilton1989,
  author  = {James D. Hamilton},
  title   = {A New Approach to the Economic Analysis of Nonstationary Time Series and the Business Cycle},
  journal = {Econometrica},
  year    = {1989},
  volume  = {57},
  number  = {2},
  pages   = {357--384}
}

@article{AAGAARD2022101335,
title = {Why capacity market prices are too high},
journal = {Utilities Policy},
volume = {75},
pages = {101335},
year = {2022},
issn = {0957-1787},
doi = {https://doi.org/10.1016/j.jup.2022.101335},
author = {Todd Aagaard and Andrew Kleit}
}

@book{creti2019economics,
  title={Economics of electricity: Markets, competition and rules},
  author={Cret{\`\i}, Anna and Fontini, Fulvio},
  year={2019},
  publisher={Cambridge University Press}
}

@book{stoft2002power,
  title={Power system economics: designing markets for electricity},
  author={Stoft, Steven},
  volume={468},
  year={2002},
  publisher={IEEE press Piscataway}
}

@article{bothwell2017crediting,
  title={Crediting wind and solar renewables in electricity capacity markets: the effects of alternative definitions upon market efficiency},
  author={Bothwell, Cynthia and Hobbs, Benjamin F},
  journal={The Energy Journal},
  volume={38},
  number={},
  pages={173--188},
  year={2017},
  publisher={SAGE Publications Sage CA: Los Angeles, CA}
}

@book{morales2013integrating,
  title={Integrating renewables in electricity markets: operational problems},
  author={Morales, Juan M and Conejo, Antonio J and Madsen, Henrik and Pinson, Pierre and Zugno, Marco},
  volume={205},
  year={2013},
  publisher={Springer Science \& Business Media}
}

@article{hach2016capacity,
  title={Capacity market design options: A dynamic capacity investment model and a GB case study},
  author={Hach, Daniel and Chyong, Chi Kong and Spinler, Stefan},
  journal={European Journal of Operational Research},
  volume={249},
  number={2},
  pages={691--705},
  year={2016},
  publisher={Elsevier}
}

@book{kirschen2018fundamentals,
  title={Fundamentals of power system economics},
  author={Kirschen, Daniel S and Strbac, Goran},
  year={2018},
  publisher={John Wiley \& Sons}
}

@article{conejo2016investment,
  title={Investment in electricity generation and transmission},
  author={Conejo, Antonio J and Baringo, Luis and Kazempour, S Jalal and Siddiqui, Afzal S},
  journal={Cham Zug, Switzerland: Springer International Publishing},
  volume={119},
  year={2016},
  publisher={Springer}
}

@article{chattopadhyay2015capacity,
  title={Capacity and energy-only markets under high renewable penetration},
  author={Chattopadhyay, Deb and Alpcan, Tansu},
  journal={IEEE Transactions on Power Systems},
  volume={31},
  number={3},
  pages={1692--1702},
  year={2015},
  publisher={IEEE}
}

@article{bhagwat2017effectiveness,
  title={The effectiveness of capacity markets in the presence of a high portfolio share of renewable energy sources},
  author={Bhagwat, Pradyumna C and Iychettira, Kaveri K and Richstein, J{\"o}rn C and Chappin, Emile JL and De Vries, Laurens J},
  journal={Utilities policy},
  volume={48},
  pages={76--91},
  year={2017},
  publisher={Elsevier}
}

@INPROCEEDINGS{IEMRE,
  author={Lanka, Vishnu Vardhan Sai and Roy, Millend and Suman, Shikhar and Prajapati, Shivam},
  booktitle={2021 Innovations in Energy Management and Renewable Resources(52042)}, 
  title={Renewable Energy and Demand Forecasting in an Integrated Smart Grid}, 
  year={2021},
  volume={},
  number={},
  pages={},
  doi={10.1109/IEMRE52042.2021.9386524}}

@article{pricing_mean_reverting,
  title={Pricing in electricity markets: a mean reverting jump diffusion model with seasonality},
  author={Cartea, Alvaro and Figueroa, Marcelo G},
  journal={Applied Mathematical Finance},
  volume={12},
  number={4},
  pages={313--335},
  year={2005},
  publisher={Taylor \& Francis}
}

@article{capacity_market_carmton,
    author = {Carmton, Peter and Stoft, Steven} ,
    title = {A Capacity Market that Makes Sense.},
    journal = {The Electricity Journal, Vol. 18, No. 7,},
    year = {2005},
}

@techreport{nyiso_capacity_market,
    title={Installed Capacity Manual. Version 11.1},
    author="NYISO capacity market products",
    institution={NYISO, New York Independent System Operator},
    year={2024},
    URL={https://www.nyiso.com/documents/20142/2923301/icap_mnl.pdf/234db95c-9a91-66fe-7306-2900ef905338}
}

@techreport{CEER2019_Portugal,
  author       = {{Council of European Energy Regulators (CEER)}},
  title        = {Annual Report on the Electricity and Natural Gas Markets in Portugal 2019},
  institution  = {Council of European Energy Regulators},
  year         = {2019},
  type         = {Tech.\ Report},
  url          = {https://www.ceer.eu/wp-content/uploads/2024/04/C20_NR_Portugal_EN.pdf},
}

@techreport{IEEFA2016_SpainCapacity,
  author       = {International Energy Finance Advisory (IEEFA)},
  title        = {Spain's Capacity Market: Energy Security or Subsidy?},
  year         = {2016},
  institution  = {IEEFA},
  note         = {Analyzes dependency of generators on capacity payments; Spain first introduced payments in 1997}
}

@techreport{MISO2023_BPM011,
  author       = {{Midcontinent Independent System Operator (MISO)}},
  title        = {Business Practice Manual BPM-011: Resource Adequacy Standards (Redline r29, August 22–23, 2023)},
  institution  = {MISO},
  year         = {2023},
  month        = aug,
  url          = {https://cdn.misoenergy.org/WPPI%20Feedback%20on%20RASC%20BPM-011-r29%20(2023-0822-23)%20Redline630187.pdf},
}

@techreport{CAISO2024_Section43A,
  author       = {{California ISO (CAISO)}},
  title        = {Fifth Replacement Electronic Tariff: Section 43A – Capacity Procurement Mechanism (June 1, 2024)},
  institution  = {California ISO},
  year         = {2024},
  month        = jun,
  url          = {https://www.caiso.com/documents/section-43a-capacity-procurement-mechanism-as-of-jun-1-2024.pdf},
}

@techreport{Svk2023_CapacityMechanism,
  author       = {{Svenska kraftnät}},
  title        = {A Future Capacity Mechanism to Ensure Resource Adequacy in the Electricity Market},
  institution  = {Svenska kraftnät},
  year         = {2023},
  month        = {},
  day          = {},
}

@article{bollerslev1986generalized,
  title={Generalized autoregressive conditional heteroskedasticity},
  author={Bollerslev, Tim},
  journal={Journal of econometrics},
  volume={31},
  number={3},
  pages={307--327},
  year={1986},
  publisher={Elsevier}
}

@misc{PJMFactSheet2025,
  author       = {{PJM Interconnection}},
  title        = {PJM Capacity Market: Promoting Future Reliability},
  howpublished = {Fact Sheet},
  year         = {2025},
  note         = {Describes PJM’s Reliability Pricing Model capacity market},
  url          = {https://www.pjm.com/-/media/about-pjm/newsroom/fact-sheets/pjm-capacity-market-promoting-future-reliability-fact-sheet.pdf},
}

@techreport{ISO_NE_IMMU_2008,
  author       = {{ISO New England Independent Market Monitor Unit}},
  title        = {2008 Annual Markets Report},
  institution  = {ISO New England Inc.},
  year         = {2008},
  howpublished = {Technical Report},
  url          = {https://www.iso-ne.com/static-assets/documents/markets/mktmonmit/rpts/ind_mkt_advsr/isone_2008_immu_report_final.pdf},
}

@article{ANDREIS2020104705,
title = {Pricing reliability options under different electricity price regimes},
journal = {Energy Economics},
volume = {87},
pages = {104705},
year = {2020},
issn = {0140-9883},
doi = {https://doi.org/10.1016/j.eneco.2020.104705},
author = {Luisa Andreis and Maria Flora and Fulvio Fontini and Tiziano Vargiolu},
}

@article{uhlenbeck1930theory,
  title={On the theory of the Brownian motion},
  author={Uhlenbeck, George E and Ornstein, Leonard S},
  journal={Physical review},
  volume={36},
  number={5},
  pages={823},
  year={1930},
  publisher={APS}
}

@book{duffie2011dark,
  title={Dark markets: Asset pricing and information transmission in over-the-counter markets},
  author={Duffie, Darrell},
  year={2011},
  publisher={Princeton University Press}
}

@misc{RSE_CapacityMarket_2024,
  author       = {{RSE (Ricerca sul Sistema Energetico)}},
  title        = {Overview of Italy Capacity Market Design (English Version)},
  year         = {2024},
  note         = {English summary document provided by RSE},
  howpublished = {\url{https://www.rse-web.it/wp-content/uploads/2024/02/01_capacityMarket-inglese.pdf}}
}

@book{mcneil2015quantitative,
  title={Quantitative risk management: concepts, techniques and tools-revised edition},
  author={McNeil, Alexander J and Frey, R{\"u}diger and Embrechts, Paul},
  year={2015},
  publisher={Princeton university press}
}

@article{PeroldSharpe1995,
  author  = {Perold, Andr{\'e} F. and Sharpe, William F.},
  title   = {Dynamic Strategies for Asset Allocation},
  journal = {Financial Analysts Journal},
  year    = {1995},
  volume  = {51},
  number  = {1},
  pages   = {149--160},
  doi     = {10.2469/faj.v51.n1.1871},
  note    = {Reprinted from FAJ 44(1): 16--27 (Jan--Feb 1988).}
}

\end{document}